\documentclass[11pt,letterpaper]{amsart}
\RequirePackage{pdf14}
\RequirePackage{fix-cm}
\usepackage[left=2.5cm,right=2.5cm,top=3.5cm]{geometry}
\ifdefined\pdfsuppresswarningpagegroup\pdfsuppresswarningpagegroup=1\fi
\usepackage{amsmath,amssymb,amsthm,mathtools}
\usepackage{booktabs,array}
\usepackage{parskip}
\usepackage{xcolor}
\definecolor{e-mail}{rgb}{0,.40,.80}
\definecolor{reference}{rgb}{.20,.60,.22}
\definecolor{citation}{rgb}{0,.40,.80}
\usepackage[colorlinks=true,linkcolor=reference,citecolor=citation,urlcolor=e-mail]{hyperref}
\usepackage[capitalise]{cleveref}
\hypersetup{pdftitle={Asymmetric multiscalar fixed points from equiangular tight frames},pdfauthor={Christopher Beem}}
\numberwithin{equation}{section}
\numberwithin{figure}{section}
\numberwithin{table}{section}
\newtheorem{thm}{Theorem}[section]
\newtheorem{prop}[thm]{Proposition}
\newtheorem{lem}[thm]{Lemma}
\newcommand{\IR}{\mathbb R}

\newcommand{\id}{\mathbf 1}

\newcommand{\ceq}{\coloneqq}
\newcommand{\rmO}{\mathrm O}
\newcommand{\Sym}{\operatorname{Sym}}
\newcommand{\Stab}{\operatorname{Stab}}
\newcommand{\rank}{\operatorname{rank}}

\newcommand{\diag}{\operatorname{diag}}

\newcommand{\MM}{\mathcal M}
\newcommand{\BB}{\mathcal B}
\newcommand{\PP}{\mathcal P}

\newcommand{\ie}{\emph{i.e.}}
\newcommand{\eg}{\emph{e.g.}}
\newcommand{\cf}{\emph{cf.}}

\title[Asymmetric multiscalar fixed points]{Asymmetric multiscalar fixed points from equiangular tight frames}
\author[C. Beem]{Christopher Beem}
\address{Mathematical Institute, University of Oxford, Woodstock Road, Oxford, OX2 6GG, UK}
\email{beem@maths.ox.ac.uk}
\address{School of Natural Sciences, Institute for Advanced Study, 1 Einstein Drive, Princeton, NJ, USA}
\email{beem@ias.edu}

\date{16 October 2026}

\begin{document}
\begin{abstract}
In their work on multiscalar RG flows, Rychkov and Stergiou posed the question of whether there exist real fixed points of the one-loop renormalization group for scalar field theories in $d=4-\epsilon$ which respect no extra symmetry beyond the universal $\mathbb{Z}_2$ field inversion. In this note, we report on a construction of a large number of such fixed points in the theory of $22001$ scalar fields, thus resolving the question. The construction uses the geometry of an \emph{equiangular tight frame} in order to reduce the (up to two-loop) fixed point equation to a two-dimensional space of couplings without assuming any extra symmetry. The main technical hurdle is then to establish the existence of such a frame whose corresponding lines preserve no symmetry. This can be achieved using the Steiner frame construction of Fickus, Mixon, and Tremain.
\end{abstract}
\maketitle
\tableofcontents
\setlength{\parskip}{10pt}

\section{\label{sec:intro}Introduction}

Consider the Wilson--Fisher RG equation for $N$ real scalar fields in $d=4-\epsilon$ dimensions with quartic potential,
\begin{equation*}
    V(\phi) = \frac{1}{4!}\lambda_{ijkl}\phi_i\phi_j\phi_k\phi_l~,
\end{equation*}
where $\lambda\in \Sym^4\mathbb{R}^N$ is the coupling tensor. The one-loop beta function takes the well-known form
\begin{equation*}
    \beta(\lambda)=-\epsilon\lambda+\frac{1}{16\pi^2}Q(\lambda)~,
\end{equation*}
where $Q:\Sym^4(\IR^N)\to\Sym^4(\IR^N)$ is a quadratic map given in coordinates by
\begin{equation*}
    Q(T)_{ijkl}=T_{ijab}T_{abkl}+T_{ikab}T_{abjl}+T_{ilab}T_{abjk}~.
\end{equation*}
Setting $\lambda = 16\pi^2\epsilon\,T$, the one-loop RG fixed point equation becomes the fixed point condition for the $Q$ operator,
\begin{equation}\label{eq:one-loop-fixed-point-eq}
    Q(T) = T~.
\end{equation}
Solutions to these one-loop RG equations have been studied since the 1970s \cite{Wilson:1971dc,Aharony:1973zz,Brezin:1973jt,Zia:1975ha} with interest continuing to the present day (see, \eg, \cite{Osborn:2017ucf,Rychkov:2018vya,Codello:2020lta,Osborn:2020cnf,Rong:2023xhz}). By and large, the study of such fixed points has been organized by symmetry principles \cite{Michel:1983in}, and often the analysis involves specifying a subgroup $G\subseteq\rmO(N)$ to be preserved by the fixed points. This serves to reduce the space of coupling tensors to those which are $G$-invariant, making computations analytically tractable.

Since the potential is quartic, any solution to \eqref{eq:one-loop-fixed-point-eq} will preserve the overall $\mathbb{Z}_2$ field inversion $\phi\to-\phi$. To date, however, all solutions with $N\geqslant 2$ that have been identified (including those found in searches with no symmetry constraints at small $N$ \cite{Codello:2020lta,Osborn:2020cnf,Rong:2023xhz}) have had a larger symmetry group. This state of affairs led the authors of \cite{Rychkov:2018vya} to articulate the specific question of whether there might exist (real) fixed points with $G=\mathbb{Z}_2$, or whether all fixed points must have strictly larger symmetry. For brevity, we will refer to fixed points with $G=\mathbb{Z}_2$ as \emph{asymmetric fixed points}.

In this paper we provide an answer to this question by constructing examples of asymmetric fixed points for an absurdly large number of scalar fields: $N=22001$. Naively, searching for fixed points with no symmetry at extremely large values of $N$ seems an unlikely strategy, since the space of couplings gets out of hand quickly and there is no organizing principle to simplify things. This conclusion turns out to be too fast. The trick exploited in this paper is to consider the two-dimensional space of potentials of the form
\begin{equation}\label{eq:potential_intro}
    V(\phi) = \frac{1}{4!}\left(\tau(\phi^2)^2 + \sigma \sum_{a=1}^M (v_a\cdot\phi)^4\right)~,
\end{equation}
where $\{v_a\}_{a=1}^M$ is a collection of unit vectors in $\IR^N$ and $(\tau,\sigma)$ are real couplings. This family of potentials will be shown to be closed under one-loop (and in fact two-loop) renormalization group flow when the unit vectors $\{v_a\}$ form what is called an \emph{equiangular tight frame}. This means that their second moment is isotropic and the angle between any pair of frame lines is the same. The most familiar example of an equiangular tight frame is the set of vectors defining a regular $N$-simplex, in which case the form of the potential in \eqref{eq:potential_intro} is the one relevant for hypertetrahedral fixed points \cite{Zia:1975ha}. However, the tight-frame condition itself does not \emph{a priori} require the arrangement of lines to have any symmetry at all beyond the overall inversion, and indeed we will see that one can construct examples with no additional symmetry, ultimately leading to asymmetric one-loop fixed points.

The line of reasoning will proceed as follows. First, we identify the range of frame parameters (the number $M$ of frame vectors and the number $N$ of scalar fields) for which the one-loop fixed point equations have real solutions and for which the quartic tensor built from the frame completely determines the corresponding set of frame lines. This latter point ensures that an asymmetric arrangement of lines will guarantee an asymmetric interaction. We then use a construction of Fickus, Mixon, and Tremain \cite{FickusMixonTremain2012} to produce frames with the required parameters using a combinatorial datum known as a \emph{Steiner system}. We first construct an auxiliary frame by embedding regular simplices in certain minimally overlapping coordinate subspaces of a fixed high-dimensional Euclidean space. For the systems relevant in this paper, this construction has a huge amount of discrete freedom associated to the choice of embeddings of each simplex, and the most technical step is to show that for almost all of these choices the set of frame lines preserves no orthogonal symmetry except overall inversion. We then take the \emph{Naimark complement} of this frame, which returns a second frame with the same number of vectors in a larger field space, but with the \emph{same} (trivial) symmetry. It will be this complementary frame that has parameters in the allowed range to guarantee an asymmetric fixed point.

The fixed points that result from the construction are real, fully interacting, and have minimal symmetry. They are potential-stable,\!\footnote{This is the terminology of \cite{Rychkov:2018vya}, meaning the quartic potential is nonnegative. It is shown \emph{loc. cit.} that any real fixed point is potential-stable.} but unstable in the RG sense to a very general class of quartic perturbations. We show that they can be improved to two-loop fixed points, but we do not establish their persistence to higher orders in the $\epsilon$ expansion (let alone a continuation to $\epsilon=1$). Finally, our existence proof does not produce any particular asymmetric frame, though if one is chosen randomly from the set of possibilities it is overwhelmingly likely to be asymmetric.

The organization of the paper is as follows. In Section~\ref{sec:mechanism} we explain the fixed point mechanism and derive sufficient criteria for an asymmetric equiangular tight frame to give an asymmetric one-loop fixed point. We also comment on RG stability and on the extension of the solutions to higher loop order. In Section~\ref{sec:construction} we describe the specific frame construction in question and state several combinatorial/geometric results that guarantee the existence of a suitable asymmetric frame. In Section~\ref{sec:discussion} we comment on some possible improvements and suggest directions for further investigation. Technical proofs related to the construction of the asymmetric frame, which are not particularly illuminating from a quantum field theory perspective, are relegated to the appendices. In Appendix~\ref{app:frames} we derive a sufficient condition for the quartic coupling to have the same symmetries as the corresponding system of frame lines. In Appendix~\ref{app:hadamard} we describe the Hadamard simplices used in our Steiner frame construction and establish some of their relevant properties. In Appendix~\ref{app:asymmetry} we prove several combinatorial results that allow us to deduce the existence of asymmetric members of the class of frames in question. In Appendix~\ref{app:smaller} we record some details of two potential examples with fewer fields, whose existence is conditional upon the (as yet unestablished) existence of certain Steiner systems.

{\bf Statement of AI use.} The construction reported in this paper is based on one originally generated by \texttt{GPT-6 Astra} at the author's prompting. The author had attempted periodically to utilize publicly available AI models to address this problem (usually through increasingly elaborate search strategies at low values of $N$), and this was the first time the effort met with any success. The first construction by the model turned out to be incomplete (it was based on one of the as-yet unestablished Steiner $2$-designs mentioned in Appendix \ref{app:smaller}) and allowed for a variety of simplifications. The proof was also significantly more complicated than necessary. The construction and proof were then refined through many rounds of iteration between the author, \texttt{GPT-6 Astra}, and \texttt{Claude Fable 5.1} until arriving at the version reported in this paper. This manuscript was written by the author with feedback from both systems and represents a distillation, simplification, and clarification (I hope) of the construction and arguments produced by them.

\section{\label{sec:mechanism}Two-coupling RG flows from equiangular tight frames}

A real equiangular tight frame is a collection $\{v_a\}_{a=1}^M$ of unit vectors spanning distinct lines in $\IR^N$ (the \emph{frame lines}) and obeying
\begin{equation}\label{eq:frame}
    \sum_{a=1}^M v_a\otimes v_a=\rho\,\id_N~,
    \qquad |v_a\cdot v_b|=\mu\quad{\rm for}\quad (a\neq b)~.
\end{equation}
The first identity is the condition for tightness of the frame, which can be rephrased as the requirement that the second moment of the set of frame vectors is isotropic. Here the number $\rho$ is known as the \emph{redundancy}, and by taking the trace of the first relation in \eqref{eq:frame} one finds that it is given by $\rho=M/N\geqslant 1$. The second condition is equiangularity, and $\mu$ is called the \emph{coherence}. Again, this parameter is actually fixed in terms of $M$ and $N$,
\begin{equation*}
    \mu^2=\frac{M-N}{N(M-1)}~.
\end{equation*}
The case $M=N$ is an orthonormal basis, while $M=N+1$ gives a regular $N$-simplex (\ie, a hypertetrahedron) after suitable sign choices. See \cite{Waldron2018} for a pedagogical introduction to tight frames.

The two tensors appearing in the ansatz \eqref{eq:potential_intro} are
\begin{equation*}
    E_{ijkl}=\delta_{ij}\delta_{kl}+\delta_{ik}\delta_{jl}+\delta_{il}\delta_{jk}~,
    \qquad F=\sum_a v_a^{\otimes4}~,
\end{equation*}
and we write the rescaled coupling tensor as $T=tE+sF$, so that $\lambda=16\pi^2\epsilon(tE+sF)$ at leading order. The fact that is absolutely crucial for our purposes is that double contractions of $F$ with $F$ and of $F$ with $E$ can be expressed again in terms of $F$ and $E$ after summing over permutations. In particular, using the relations in \eqref{eq:frame} we are directly led to the relations
\begin{equation}\label{eq:FF}
    \begin{split}
    F_{ijmn}F_{mnkl}&=\mu^2\rho^2\delta_{ij}\delta_{kl}+(1-\mu^2)F_{ijkl}~,\\[3pt]
    E_{ijmn}F_{mnkl}&=\rho\delta_{ij}\delta_{kl}+2F_{ijkl}~.
    \end{split}
\end{equation}
Combined with the standard relation $Q(E)=(N+8)E$, we get the following action of $Q$ on our two-coupling quartic tensor,
\begin{equation*}
    Q(tE+sF)=\left[(N+8)t^2+2\rho ts+\mu^2\rho^2s^2\right]E+\left[12ts+3(1-\mu^2)s^2\right]F~.
\end{equation*}
As long as $E$ and $F$ are independent,\!\footnote{If $E$ and $F$ are not independent, then $F$ is $\rmO(N)$-invariant, so in particular the tight-frame tensor is not asymmetric. Consequently, when we later construct an asymmetric frame tensor it will follow that $E$ and $F$ are independent.} the one-loop fixed point equations for this two-coupling system become (assuming $s\neq0$),
\begin{equation}\label{eq:scalar}
    (N+8)t^2+2\rho ts+\mu^2\rho^2s^2=t~, \qquad 12t+3(1-\mu^2)s=1~.
\end{equation}
Solving the second equation with $t=(1-3(1-\mu^2)s)/12$, one is left with a quadratic equation determining $s$, $As^2+Bs+C=0$, where
\begin{equation*}
    \begin{split}
    A&=\frac{N+8}{16}(1-\mu^2)^2-\frac\rho2(1-\mu^2)+\mu^2\rho^2~,\\[3pt]
    B&=\frac{4\rho-(N+2)(1-\mu^2)}{24}~,\\[3pt]
    C&=\frac{N-4}{144}~.
    \end{split}
\end{equation*}
The discriminant $\Delta=B^2-4AC$, when written in terms of $M$ and $N$, is given by
\begin{equation*}
    \Delta=\frac{M^2}{144N^3(M-1)^2}\left(16M^2-8M(N^2+N+2)+N(9N^2-10N+17)\right)~.
\end{equation*}
The existence of a real fixed point is thus completely determined by the values of $M$ and $N$. For $2\leqslant N\leqslant 4$ one can show that the discriminant is positive for all $M\geqslant N$, so there is always a real solution. However, in this range it turns out that all equiangular tight frames preserve additional symmetry, so we will restrict to $N\geqslant 5$ in all that follows.\!\footnote{By the Gerzon bound \cite{LemmensSeidel1973} and the integrality conditions of \cite{sustik2007existence}, the only real equiangular tight frames with $N\leqslant 4$ are orthonormal bases, regular simplices, and the six diagonals of an icosahedron in $\IR^3$, each unique up to isometry and each with a large symmetry group.} In this case, there is an interval of $M$ values for which the solutions are complex. We will be interested in the lower range of $M$ values with distinct real solutions, where the number of vectors is not dramatically larger than the dimension. This is the \emph{reality interval} (imposing $M\geqslant N$),
\begin{equation}\label{eq:window}
    N\leqslant M< M_-(N)~,\qquad M_-(N)\ceq\frac{N^2+N+2-\sqrt{(N-1)^3(N-4)}}4~.
\end{equation}
We note that for $N\geqslant 5$ one always has $M_-(N)<11N/8$.

We want to be sure that an asymmetric choice of frame (\ie, a frame whose corresponding set of lines has no symmetry beyond the universal $\mathbb{Z}_2$ inversion) implies an asymmetric one-loop fixed point. This is not automatic because even if the frame is asymmetric, it could be that at the level of its fourth moment some symmetry is restored (as is the case for the second moment). It turns out that restricting to sufficiently small redundancy is sufficient to avoid this issue. Indeed, one can prove the following:

\begin{lem}
For any unit tight frame of distinct lines with redundancy $\rho<3/2$ (not necessarily equiangular), the fourth-moment tensor determines the frame lines, so one has
\begin{equation}\label{eq:symmetry}
    \Stab_{\rmO(N)}(F)=\Stab_{\rmO(N)}\{\IR v_a\}~.
\end{equation}
\end{lem}

The idea behind the proof of this lemma is that one can regard $F$ as an operator on symmetric matrices, in which case its image is the span of the rank-one projectors onto the individual frame lines. Low redundancy then ensures that this span contains no other rank-one projectors, so $F$ encodes the original line arrangement. The proof is given in Appendix~\ref{app:frames}.

Because the reality interval for $M$ lies entirely below $3N/2$, the identification of stabilizers holds for any real fixed points in that interval.\!\footnote{This was the reason for restricting to the lower reality interval.} Furthermore, with the restrictions $N\geqslant 5$ and $\rho<3/2$ one finds that $A>0$ and $C>0$. Hence for $N\leqslant M < M_-(N)$ there are two distinct, nonzero real solutions for $s$, and since $E$ is $\rmO(N)$-invariant, $\Stab_{\rmO(N)}(tE+sF)=\Stab_{\rmO(N)}(F)$. We can then summarize the state of affairs as follows:
\begin{prop}\label{prop:criterion}
Let $N\geqslant 5$ and let $M<M_-(N)$ unit vectors form a real equiangular tight frame in $\IR^N$. If the only orthogonal symmetries of the collection of frame lines are $\pm\id$, then \eqref{eq:scalar} has two distinct real solutions, each of which gives an asymmetric one-loop fixed point.
\end{prop}
Such a fixed point will be fully interacting, as any tensor that decomposes as a direct sum is preserved by inversion in either summand.

\subsection{\label{subsec:instability}Stability under quartic perturbations}

At this level of generality we can also address the question of RG stability against relevant perturbations. It turns out that the frame relations allow us to identify a large space of relevant perturbations outside of the two-parameter ansatz.

To determine the scaling dimensions of quartic perturbations, we look for eigentensors of the total derivative of the one-loop beta function at the fixed point, which is given by $\epsilon(-\id+DQ_{T_*})$, where $T_\ast$ is the rescaled coupling tensor evaluated at a one-loop fixed point. Up to the overall factor of $\epsilon$, the eigenvalue equation is then
\begin{equation*}
    (-\id + DQ_{T_\ast})(S)_{ijkl} = -S_{ijkl} + 6(T_\ast)_{(ij|ab}S_{ab|kl)} = \kappa S_{ijkl}~,
\end{equation*}
where parentheses denote normalized symmetrization over the external indices; this is the stability-matrix eigenvalue problem of \cite{Osborn:2020cnf}, with the same normalization of $\kappa$. The one-loop RG eigenvalue is then given by $\epsilon\kappa$, and the one-loop scaling dimension of the corresponding quartic operator $\mathcal O_S=S_{ijkl}\phi_i\phi_j\phi_k\phi_l$ is therefore
\begin{equation*}
    \Delta_{\mathcal O_S}=4-\epsilon+\epsilon\kappa~,
\end{equation*}
and $\kappa<0$ means that the operator is relevant.

Now consider symmetric quartic tensors whose contraction with two copies of any frame vector vanishes,
\begin{equation*}
    \mathcal K=\left\{S\in\Sym^4(\IR^N)\,:\,S(v_a,v_a,\cdot,\cdot)=0~,\quad {a=1,\ldots,M}\right\}~.
\end{equation*}
Summing this condition over the frame vectors and using tightness shows that any such $S$ is, in particular, traceless. We also have simple contraction identities
\begin{equation*}
    E_{ijab}S_{abkl} = 2S_{ijkl}~,\qquad F_{ijab}S_{abkl} = 0~,
\end{equation*}
so the linearized beta function at $T_\ast$ preserves $\mathcal K$ and acts there by the scalar $\epsilon\kappa_{\mathcal K}=\epsilon(12t_\ast-1)$. For our one-loop fixed points with nonvanishing $s$, \eqref{eq:scalar} gives $12t_\ast-1=3(\mu^2-1)s_\ast$, so every nonzero coupling in $\mathcal K$ is relevant if $s_\ast>0$, which will be the case for both solutions in our construction.

A crude lower bound on the number of such relevant couplings is
\begin{equation*}
    \dim\mathcal K\geqslant\binom{N+3}{4}-M\binom{N+1}{2}~.
\end{equation*}
Here we have used that the vanishing condition for each frame vector imposes at most $\dim\Sym^2(\IR^N) = \binom{N+1}{2}$ linear constraints on the $\binom{N+3}{4}$-dimensional space of symmetric quartic tensors.

\subsection{Two loops and higher}

We further observe that at two loops (in, say, ${\rm MS}$ or $\overline{\rm MS}$ schemes) the beta function still preserves the specified two-dimensional space of couplings. This follows again from the now-familiar frame identities. The relevant contractions appearing in the two-loop beta function \cite{Jack:1990eb} (see \cite[Eq.~(3.6)]{Osborn:2017ucf} for the explicit form) are of the form
\begin{equation*}
T_{ijmn}T_{mkab}T_{nlab}~,\qquad \gamma^{(2)}_{ia}(T)T_{ajkl}~,
\end{equation*}
along with various index permutations. Here the two-loop field renormalization is given by $\gamma^{(2)}_{ij}(T)=T_{iabc}T_{jabc}/12$. The vertex terms reduce for the same reason as at one loop: a pair of $F$ tensors joined by two contracted indices simplifies by \eqref{eq:FF}. In particular,
\begin{equation*}
    F_{ijmn}F_{mkab}F_{nlab}
      =\mu^2(1-\mu^2)\rho^2\delta_{ij}\delta_{kl}
       +\bigl(\mu^2\rho^2+(1-\mu^2)^2\bigr)F_{ijkl}~.
\end{equation*}
Terms containing any $E$ insertions reduce upon resolving each $E$ into delta functions and using the same identities, together with $F_{ijaa}=\rho\delta_{ij}$. The field renormalization is also isotropic, $\gamma^{(2)}_{ij}(tE+sF)\propto \delta_{ij}$. Thus the full two-loop beta function is again a sum of $E$ and $F$ (with some cubic polynomials in $t$ and $s$ as coefficients).

The extension of the fixed points to two loops now follows on general grounds. In the plane spanned by $E$ and $F$ the fixed point condition (with a factor of $s$ restored) is a pair of quadratic equations in $(t,s)$ perturbed by a cubic two-loop part suppressed by a further power of $\epsilon$. A solution of the one-loop equations at which the linearization of that system is invertible is therefore shifted at relative order $\epsilon$ and not destroyed. That invertibility holds at both of the solutions to \eqref{eq:scalar} can be understood by noting that there are four distinct solutions to the full system of two quadratics (the free theory, the $\rmO(N)$ fixed point, and the two roots with $s\neq0$), so, by B\'ezout's theorem, each is a simple zero.

At three loops, the field renormalization introduces no obstruction to closure on the two-coupling subspace. In the same normalization as above, it is given by \cite{Jack:1990eb} (in the form of \cite[Eq.~(3.6)]{Osborn:2017ucf})
\begin{equation*}
    \gamma^{(3)}_{ij}(T)=-\frac1{16}T_{iklm}T_{jknp}T_{lmnp}~.
\end{equation*}
This contraction still includes double contractions, so the same sort of manipulations lead to an isotropic renormalization $\gamma^{(3)}_{ij}(tE+sF)\propto\delta_{ij}$. The contribution to the quartic beta function is given by $\gamma^{(3)}_{ia}(T)T_{ajkl}$ together with the other three external-leg terms; the sum remains in ${\rm span}\{E,F\}$.

The new structure at this order instead comes from a vertex correction where four quartic vertices are contracted pairwise (giving a diagram in the shape of a tetrahedron). This
contributes to the beta function a term of the form
\begin{equation*}
    H(T)_{ijkl}=T_{iabc}T_{jade}T_{kbdf}T_{lcef}~.
\end{equation*}
All other three-loop vertex contractions contain pairs of vertices with at least two contracted indices between them and reduce as at two loops. The tetrahedron-shaped contractions involving any $E$ vertices also reduce to sums of $E$ and $F$. But for the tetrahedral contraction of $F$ vertices one obtains a nontrivial new condition for closure on the two-coupling subspace,
\begin{equation*}
    H(F)\in\mathrm{span}\{E,F\}~.
\end{equation*}
This is a tight-frame version of the additional condition in \cite[Eq.~(5.27)]{Osborn:2017ucf}. In terms of frame vectors, this tensor takes the form
\begin{equation*}
\sum_{a,b,c,d}(v_a\cdot v_b)(v_a\cdot v_c)(v_a\cdot v_d)(v_b\cdot v_c)(v_b\cdot v_d)(v_c\cdot v_d)\,v_a\otimes v_b\otimes v_c\otimes v_d~.
\end{equation*}
There is no automatic simplification of this expression coming from the frame conditions, so generically at three loops we expect closure on the two-coupling space to fail. When this is the case, the fate of the fixed point is uncertain.

\section{\label{sec:construction}Asymmetric frames from overlapping simplices}

Having reduced the problem to one of producing a suitable frame, we describe in this section a construction which is guaranteed to produce one with the required dimensionality parameters and with no extra symmetries. This uses a construction of Fickus, Mixon, and Tremain \cite{FickusMixonTremain2012} (for real frames an equivalent construction had already been given, in the language of strongly regular graphs, by Goethals and Seidel in \cite{GoethalsSeidel1970}). What will be important for us is that the construction allows a huge number of independent choices such that for some of those choices the frame will necessarily be asymmetric. We describe below the geometric construction and summarize the counting arguments that establish asymmetry, but we will relegate many details to appendices.

\subsection{The geometric construction}

A Steiner $2$-design, or Steiner system $S(2,k,v)$, is a set $\PP$ of $v$ points (called \emph{design points}) along with a collection $\BB$ of $k$-point subsets of $\PP$ (called blocks) such that every pair of design points belongs to exactly one block \cite[Part~II]{ColbournDinitz2007}. The blocks through a given point partition the other $v-1$ points into groups of $k-1$, so the number of blocks containing a given point is $r=(v-1)/(k-1)$. One then has a total of $D=vr/k$ blocks. The existence of an $S(2,k,v)$ system places highly nontrivial constraints on $k$ and $v$; see \cite[\S II.3]{ColbournDinitz2007} for the status at small $k$.

The construction of \cite{FickusMixonTremain2012} uses this combinatorial data (and a bit more, described below) to construct a family of equiangular tight frames of $M=v(r+1)$ lines in $\IR^D$. One first assigns to each block $B_i$ a distinct coordinate vector $e_{B_i}\in\IR^D$. The corresponding axis will be referred to as a \emph{block axis}. A design point $x$ then specifies the $r$-dimensional coordinate subspace $W_x\subset\IR^D$ spanned by the axes associated to the blocks containing it. The condition that each pair of design points belongs to exactly one block means that any two such coordinate subspaces share exactly one coordinate axis. In turn, each axis belongs to exactly $k$ subspaces. Thus the Steiner system provides a collection of minimally (but nontrivially) overlapping coordinate subspaces of $\IR^D$ with the further condition that every axis belongs to the same number of such subspaces.

The corresponding frames are constructed by placing, for each design point $x$, a regular $r$-simplex in the block subspace $W_x$. We require that these simplices be positioned in a very specific way relative to the coordinate axes: each vertex must have coordinate $\pm1/\sqrt r$ along each block axis of $W_x$. A regular simplex of this type is equivalently determined by a Hadamard matrix of order $h=r+1$ with one row taken to be all positive: its orthogonal columns have entries $\pm1$, and deleting the all-positive row gives the vertices in $\IR^r$ as the column vectors after rescaling by $1/\sqrt r$. We will refer to such a simplex as a Hadamard simplex. Thus the frame construction requires a choice of a Hadamard simplex for each design point.

Assuming we have chosen a particular model Hadamard simplex for a given design point, there is freedom in how the model simplex is embedded into the corresponding coordinate subspace. This freedom amounts to choosing how to match the (oriented) axes of the model simplex to the block axes through the design point. There are $2^r r!$ such \emph{signed matchings}, and they can be chosen independently at each design point.

Within a given simplex the inner product of two distinct vertices is $-1/r$; for pairs of vertices coming from different simplices the inner product has magnitude $1/r$ because the corresponding subspaces share only one axis. Hence the frame is equiangular. Furthermore, each simplex is a tight frame for its own subspace, and every axis belongs to the same number of subspaces, so the union is tight. Thus the construction produces an equiangular tight frame of $M=v(r+1)$ lines in $\IR^D$.

To attain sufficiently low redundancy, it will be necessary to further pass to the \emph{Naimark complement} of the Steiner frame; the explicit construction is given in, \eg, \cite[Section~2.2]{FickusJasperKingMixon2018}. Geometrically, the original frame and its complement are projections of a common orthonormal basis of $\IR^M$ onto complementary subspaces, with the projected vectors normalized to unit length. The complement is again an equiangular tight frame, now of $M$ vectors in $\IR^{M-D}$, and is unique up to orthogonal transformation. Starting from a Steiner frame, the complementary frame therefore has parameters
\begin{equation*}
    N=v(r+1)-\frac{vr}{k}~,
    \qquad M=v(r+1)~,
    \qquad \frac MN=1+\frac{r}{k(r+1)-r}~.
\end{equation*}
Crucially, since both frames consist of distinct lines spanning their spaces, the orthogonal stabilizer of either set of lines is faithfully represented by the signed permutations of the index set preserving the Gram matrix, and these coincide for a frame and its Naimark complement \cite[Theorem~9.2]{Waldron2018}. Thus an asymmetric Steiner frame will give an asymmetric complementary frame. For us, it will be these complementary frames that have sufficiently small redundancy to give rise to an asymmetric fixed point. However, the symmetry analysis will be performed using the original frame where the geometry is more transparent.

\subsection{Avoiding symmetries in a specific construction}

Our specific argument uses an $S(2,9,441)$ system. Such a system is obtained from the \emph{group divisible design} $\mathrm{GD}[9,1,9;441]$ of \cite[Theorem~4.6]{Greig1998} (the case $t=6$, $q=8t+1=49$), which has $49$ groups of size $9$; adjoining the groups as additional blocks yields the Steiner system \cite[\S2]{Greig1998}. We will therefore have $M=24696$ frame lines in $D=2695$ dimensions, with coordinate subspaces of dimension $r=55$. We then need Hadamard matrices of order $h=56$ to define the Hadamard simplex at each of the $441$ design points. In particular, we will use Hadamard matrices that are realized by doubling Hadamard matrices of order $28$, as described in Appendix~\ref{app:hadamard}. There are $487$ inequivalent Hadamard matrices of order $28$ \cite{Kimura1994},\!\footnote{Two Hadamard matrices are equivalent when one is obtained from the other by permuting and negating rows and columns.} and inequivalent matrices have inequivalent doubles (see Appendix~\ref{subapp:doubled_simplex}), so we can assign pairwise inequivalent Hadamard simplices to the $441$ design points.

The discrete freedom in the frame construction is the choice of signed matching at each design point, giving us a set $\MM$ of constructions with
\begin{equation*}
    |\MM|=(2^{55}55!)^{441}\approx 1.6 \cdot 10^{39540}~.
\end{equation*}
We will not need to assume that different matchings give inequivalent frames; we will instead proceed by restricting the choices that can be made from this finite set.

It turns out that the vast majority of this enormous number of matching choices is guaranteed to yield frames with no orthogonal symmetry beyond $\pm\id$. To show this, we first have to deal with the fact that an orthogonal symmetry of the frame lines does not \emph{a priori} respect the original grouping of lines into simplices, whereas an analysis of symmetries that preserve simplices is more straightforward. To dispense with this complication, we show that for most choices of signed matchings the only possible symmetries preserve the groupings of lines into the original design point simplices. Indeed, we have the following:

\begin{lem}\label{lem:simplex_recovery}
For all but a fraction smaller than $1/572$ of the set of signed matchings in $\MM$, the arrangement of the frame lines into groups specified by the original design point simplices is determined uniquely by the set of lines.
\end{lem}

This is proved in Appendices~\ref{subapp:dependencies} and~\ref{subapp:extra_simplices}. The idea behind the proof is that the original simplices can be recovered by finding the shortest linear dependencies amongst the frame lines. There are also always dependencies of this same size involving lines from pairs of the original simplices, but these are unambiguously distinguished from the original simplices upon counting the number of shortest dependencies involving each pair of frame lines. If there are shortest linear dependencies involving lines from more than two of the original simplices, this complicates the situation, but such cases can be avoided as the number of signed matchings giving rise to such additional dependencies is bounded, leading to Lemma \ref{lem:simplex_recovery}.

When the original simplices are recoverable, any orthogonal symmetry of the frame lines must permute their spans. Such a symmetry also permutes the pairwise intersections of the spans, \ie, the block axes. If it mapped $W_x$ to $W_y$ with $y\neq x$, its restriction would be a signed permutation of block axes carrying the lines of the Hadamard simplex at $x$ to those at $y$, which, after restoring the constant row, would be an equivalence of the two Hadamard matrices. Since we chose these to be pairwise inequivalent, the symmetry fixes every span, and hence every shared axis. Thus any such symmetry must act as a collection of reflections of the block axes.

It remains to bound the number of matchings that are compatible with each such symmetry and sum over (candidate) symmetries. This is done in Appendix \ref{app:asymmetry}. Combined with the bound in Lemma \ref{lem:simplex_recovery}, the result is the following overall bound on the number of signed matchings with extra symmetries,
\begin{equation}\label{eq:bad_fraction}
    \frac{\#\{m\in\MM:\text{the resulting frame has a symmetry beyond }\pm\id\}}
         {|\MM|}<\frac{1}{572}~.
\end{equation}
In particular, asymmetric frames exist, and most choices of signed matchings will produce one.

We can also give a crude lower bound on the number of inequivalent frames obtained this way. Among the asymmetric constructions retained by the argument above, an equivalence must fix every block axis, giving at most $2^{2694}$ possible ambient transformations up to overall inversion. For each such transformation, the possible redundancy in the choice of matchings is bounded by $55!$ axis permutations at each design point and, for each permutation, at most four compatible sign choices (associated to the diagonal symmetries of the model simplex lines, \cf\ Appendix~\ref{subapp:doubled_simplex}). One equivalence class can thus account for at most $2^{2694}(4\cdot55!)^{441}$ matching choices. Dividing the more than $571|\MM|/572$ asymmetric choices by this upper bound gives more than $10^{6224}$ inequivalent asymmetric frames.

\subsection{The asymmetric fixed points}

Returning to the fixed point equations, the Naimark complement of any asymmetric frame obtained from the above strategy will have parameters
\begin{equation*}
    M=24696~,\qquad N=24696-2695=22001~,\qquad \rho=\frac{504}{449}~, \qquad \mu=\frac{1}{449}~.
\end{equation*}
These parameters satisfy \eqref{eq:window}, and thus give rise to asymmetric one-loop fixed points. Letting $F$ be the tensor of the complementary frame, the two fixed points are $T_\pm=t_\pm E+s_\pm F$, where
\begin{equation*}
    t_\pm=\frac{15101\pm200\sqrt{201}}{2640002412}~,
    \qquad s_\pm=\frac{201601}{604800}(1-12t_\pm)~.
\end{equation*}
These scalar couplings are the same for every signed matching.

The unstable directions identified in Section~\ref{subsec:instability} now account for the vast majority of quartic perturbations. Indeed, the relevant eigenspace $\mathcal{K}$ has dimension at least $9759126792219805$, which is more than $99.9\%$ of the space of quartic couplings.

\section{\label{sec:discussion}Discussion}

The examples above answer the one-loop existence question, but the general mechanism is perhaps the most interesting aspect of the story. A low-dimensional space of quartic couplings can be preserved by the RG flow (at low loop order) without imposing any symmetry using certain metric properties of a set of vectors in the field space. The rather large value of $N$ in our example is a consequence of the particular construction used to produce an asymmetric frame, but we see no particular reason to expect that asymmetric scalar fixed points won't exist at lower values of $N$.

Within the Steiner family, reducing $N$ would require finding designs with suitable parameters as well as Hadamard simplices for which analogous symmetry bounds work out. Two potential improvements would be $N=2737$ (requiring a $2$-$(153,9,1)$ design) and $N=1972$ (requiring a $2$-$(136,10,1)$ design). If either design exists, we expect that suitable matching choices would again give asymmetric frames; the corresponding parameters are recorded in Appendix~\ref{app:smaller}. To our knowledge neither design is known to exist; see the tables in \cite[Part~II, \S3]{ColbournDinitz2007} and, for $k=9$, \cite[Table~2]{AbelBluskovGreig2004}. It would also be interesting to explore whether there might be other constructions of equiangular tight frames that could provide asymmetric examples at lower dimensionality.

An obvious physics question is to ascertain the fate of the fixed points beyond two loops. We have not made any significant effort in this direction yet.

Finally, while our proof establishes that most matching choices succeed, we have not produced any concrete choice of frame that is verified to be asymmetric. It could be of interest to produce an exact, verifiable example.

\section*{Acknowledgments}
I would like to thank Nikolay Bobev and Yifan Wang for helpful conversations related to this work. I gratefully acknowledge the support of the J. Robert Oppenheimer Visiting Professor Endowment at the Institute for Advanced Study. My work was also partially supported by STFC grant ST/X000761/1.

\bigskip

\appendix

\section{\label{app:frames}Recovering frame lines from the coupling tensor}

In this appendix we prove the identity \eqref{eq:symmetry},
\begin{equation*}
    \Stab_{\rmO(N)}(F)=\Stab_{\rmO(N)}\{\IR v_a\}~,
\end{equation*}
for unit tight frames of distinct lines with redundancy $\rho<3/2$. Equiangularity is not needed for this analysis. The idea is to recover the span of the frame line projectors from $F$, and then to show that it contains no other rank-one projectors. Of course the frame lines also determine $F$, so this reconstruction identifies their stabilizers.

Let $\Pi_a$ be the projector onto the frame line $\IR v_a$, and let ${\mathcal S}=\mathrm{span}\{\Pi_a\}\subset\Sym^2(\IR^N)$. $F$ can be thought of as an operator on symmetric matrices by contracting two indices,
\begin{equation*}
    F(X)=\sum_a\langle v_a|X|v_a\rangle\,\Pi_a~.
\end{equation*}
Each summand in the above is a positive semidefinite rank-one operator on $\Sym^2(\IR^N)$ (with the Hilbert--Schmidt inner product). The image of a sum of positive semidefinite operators is the span of their images, so ${\mathcal S}$ is exactly the image of $F$.

We next want to show that for sufficiently low redundancy, the only rank-one projectors in ${\mathcal S}$ are the $\{\Pi_a\}$. This will establish that the frame lines are reconstructible from $F$. Suppose a rank-one projector $X$ lies in ${\mathcal S}$, so that it can be written as
\begin{equation*}
    X=\sum_{a\in\mathcal A}\lambda_a\Pi_a~,
\end{equation*}
where all $m=|\mathcal A|$ coefficients are nonzero. Let $L=\mathrm{span}\{v_a\}_{a\in\mathcal A}$ be the span of these $m$ vectors and let $\ell=\dim L$. Let $\{e_1,\dots,e_{\ell}\}$ be an orthonormal basis of $L$ and $\Pi_L$ its orthogonal projector. We then have the following inequality,
\begin{equation*}
    m=\sum_{a\in\mathcal A}\langle v_a|\Pi_L|v_a\rangle
      \leqslant\sum_a\langle v_a|\Pi_L|v_a\rangle
      =\sum_{i=1}^{\ell}\sum_a\langle e_i|v_a\rangle\langle v_a|e_i\rangle
      =\rho \ell~.
\end{equation*}
Now let $X=V_{\mathcal A}\Lambda V_{\mathcal A}^{\dagger}$ where $V_{\mathcal A}$ is an $N\times m$ matrix whose columns are the $\{v_a\}_{a\in\mathcal A}$ and $\Lambda$ is the diagonal $m\times m$ matrix with the $\lambda_a$ entries on the diagonal. The range of $V_{\mathcal A}$ is $L$, so its rank is $\ell$. Since every $\lambda_a$ is nonzero, $\Lambda V_{\mathcal A}^{\dagger}$ also has rank $\ell$. Applying the rank inequality $\rank(AB)\geqslant\rank A+\rank B-m$ to these $(N\times m)$ and $(m\times N)$ matrices, we find
\begin{equation*}
    1=\rank X\geqslant2\ell-m
      \geqslant m\left(\frac2\rho-1\right)~.
\end{equation*}
If $\rho<3/2$, this forces $m<3$. In the case $m=2$, distinctness of the frame lines fixes $\ell=2$, and the same rank inequality gives $\rank X\geqslant2$. Thus $m=1$, and $X$ is one of the original projectors. The tensor $F$ therefore determines the frame lines, and their orthogonal stabilizers agree.

\section{\label{app:hadamard}Doubled Hadamard simplices}

The counting arguments that we will use in Appendix~\ref{app:asymmetry} require us to understand two properties of our model simplices relative to their coordinate axes:
\begin{itemize}
    \item How many triples of coordinate axes support the centroids of fourteen-vertex faces;
    \item Which combinations of coordinate reflections give automorphisms of the simplex.
\end{itemize}
For a particular class of model simplices, we will show that there are at most $27$ (in fact exactly $27$) admissible triples of axes, and that there is a unique nontrivial coordinate reflection preserving their vertices.

We start with a Hadamard matrix of order $28$. Choose any such matrix $P$ normalized to have positive first row and column, and then build the doubled matrix
\begin{equation}\label{eq:doubled_hadamard}
    H=\begin{pmatrix}P&P\\P&-P\end{pmatrix}~.
\end{equation}
This $H$ is also normalized and is a Hadamard matrix of order $56$. Our model simplices are obtained from such an $H$ by deleting the constant (first) row and rescaling the resulting columns to give $56$ unit vectors in $\IR^{55}$. The arguments below apply for any such $H$, using only the general doubled structure in \eqref{eq:doubled_hadamard}.

Both the face count and the reflection argument will follow from a parity obstruction for the order-$28$ Hadamard matrix $P$: no four distinct rows or columns of $P$ can have constant entrywise (Hadamard) product. We give the argument for rows; the column case follows by transposition. Without loss of generality, we can multiply columns by signs to make one of the chosen rows constant, and negate another row if necessary so that the product is positive. Orthogonality of a second row means that, up to reordering columns, it has the form $(+^{14},-^{14})$. Orthogonality to the first two rows then forces the third, up to further reordering within each half, to have the form $(+^7,-^7,+^7,-^7)$. By our assumption of positive entrywise product, the fourth row will have the form $(+^7,-^7,-^7,+^7)$. Summing these four rows shows that any fifth row orthogonal to all four would have to be orthogonal to $(+^7,0^{21})$, which is impossible for a row with entries $\pm1$.

\subsection{\label{subapp:face_centroids}Face centroids and coordinate axes}

Below we will use $\pm1$ coordinates for the vertices (\ie, the columns of $H$ with the constant row deleted) rather than the rescaled unit vectors. These vertices then have squared norm $55$ and mutual inner products $-1$. For a face $C$ consisting of $c$ vertices with centroid $v_C$, we write $u=56/c$, and have
\begin{equation}\label{eq:face_centroid}
    v_C\cdot v_C=u-1~,\qquad v_a\cdot v_C=-1\quad{\rm for}\quad v_a\notin C~.
\end{equation}
In Appendix \ref{app:asymmetry} the faces of interest will have their centroids supported on exactly $u-1$ coordinate axes with coordinates $\pm1$, consistent with the first identity.

We will be interested in the case where $c=14$ so $u=4$. Suppose we have a face whose centroid is supported entirely on three axes with coordinates of magnitude one. Without loss of generality, assume the coordinates are positive, in which case all of the vertices in the face must have coordinates $+1$ on all three axes. Every omitted vertex has inner product $-1$ with the centroid by \eqref{eq:face_centroid}, so their coordinates on the three axes must sum to $-1$. Together with the face vertices' pattern $(+++)$, the only possible sign patterns on the three axes are therefore
\begin{equation*}
    (+,+,+)~,\qquad (+,-,-)~,\qquad (-,+,-)~,\qquad (-,-,+)~.
\end{equation*}
Their products are all positive. Thus the entrywise product of the three coordinate rows corresponding to the axes supporting the centroid must be constant.

The rows of $H$ come in pairs $(p,p)$ and $(p,-p)$, with $p$ a row of $P$. For three nonconstant rows of $H$ to have constant product, their first halves must multiply to a constant row $\pm\mathbf1$. However, by the above parity obstruction, three distinct nonconstant rows of $P$ cannot have Hadamard product equal to a constant row. Thus two of the rows of $P$ must agree and the third must be constant. The allowed triples are thus precisely
\begin{equation*}
    \{(p,p),(p,-p),(\mathbf1,-\mathbf1)\}~,
    \qquad p\text{ a nonconstant row of }P~.
\end{equation*}
There are $27$ choices of $p$, and hence at most $27$ triples of axes supporting $14$-vertex face centroids.\!\footnote{One can easily argue that any such triple of coordinate rows is the support of the centroid of some $14$-vertex face, so exactly $27$ triples of axes support such faces, but the ``at most'' result is actually sufficient for our purposes.}

\subsection{\label{subapp:doubled_simplex}Symmetries of the model simplices}

We first show that there are no nontrivial coordinate-reflection symmetries of the smaller simplex defined by $P$. Consider a nonidentity $28\times 28$ diagonal sign matrix $\Sigma$ acting on $P$ from the left. For this to be a symmetry of the corresponding simplex it must act as a permutation of the columns of $P$, and since it is an involution it must exchange the columns in pairs. (No column can be fixed because every entry of every column is nonzero and the sign matrix $\Sigma$ thus inverts at least one entry.) The entrywise product of any two paired columns is the diagonal of $\Sigma$, so the four columns in any two of the pairs have entrywise product $+1$, contradicting the parity obstruction. Thus $P$ has no nontrivial coordinate-reflection symmetries.

Now consider a nonidentity coordinate reflection $\Xi$ of $H$. As with $\Sigma$, it must act as a permutation of columns, now of $H$, that pairs all columns. This will send the all-positive first column to $(p,p)$ or $(p,-p)$ for some column $p$ of $P$. The diagonal of $\Xi$ is therefore precisely of that form. Considering the action on a column of the form $(q,\pm q)$, we see that multiplication by $p$ must permute the columns of $P$. Given the above, it follows that $p=\mathbf1$, leaving only $\id_{56}$ and $\diag(1^{28},-1^{28})$ as possible choices for $\Xi$. After deleting the constant row, there is thus a unique nonidentity reflection preserving the simplex vertices, which gives a distinguished split into $27$ fixed and $28$ reversed coordinate axes.

Considering simplex lines instead, these actions and their overall inversions are the only diagonal symmetries. Indeed, an isometry preserving the lines sends the vertices to a permutation of themselves up to an overall common sign (preserving inner products forbids independent negations of the vertices).

A consequence of the above symmetry result is that for inequivalent Hadamard matrices $P$, the corresponding doubled Hadamard matrices $H$ are also inequivalent. Indeed, up to sign, $\Xi=\diag(1^{28},-1^{28})$ is the only coordinate reflection preserving the column lines of $H$, so by restricting to eigenspaces of $\Xi$ we can recover $P$.

\section{\label{app:asymmetry}Existence of asymmetric Steiner frames}

In this appendix we establish the existence of asymmetric Steiner frames. We fix once and for all the design with $(v,k,r,h,D)=(441,9,55,56,2695)$ and $441$ inequivalent Hadamard simplices of order $56$, one assigned to each design point; only the $2^r r!$ choices of signed matching at each design point are allowed to vary. Within this family, we first show that the simplices used in the frame construction can almost always be recovered from the collection of unlabelled frame lines. We then bound the proportion of matching choices that produce frames symmetric under orthogonal transformations that fix the simplices. When all is said and done, we will see that the vast majority of signed matchings will lead to asymmetric frames.

For a subset $\mathcal A\subseteq\MM$, we write $p(\mathcal A)=|\mathcal A|/|\MM|$ for the fraction of the full set of matching choices it represents. We will require no assumption that different matchings give rise to inequivalent frames.

\subsection{\label{subapp:dependencies}Recovering the original simplices}

An orthogonal symmetry of the frame lines will not necessarily respect the grouping of those lines into the simplices used in the Steiner frame construction. However, it turns out that usually those original simplices \emph{can} be unambiguously identified from the frame lines by studying the shortest linear dependencies amongst the frame lines. When this is the case, any symmetry of the frame \emph{will} respect the original simplices.

Because distinct unit frame vectors have overlap of magnitude $1/r$, a linearly dependent set of frame lines must contain at least $r+1=h=56$ lines. Equality holds precisely when the lines can be oriented as the vertices of a regular $r$-simplex \cite[Theorem~3.1]{FickusJasperKingMixon2018}. We will call such a set of $56$ linearly dependent frame lines a \emph{short dependent set}.

Suppose a short dependent set involves frame vectors from $u>1$ of the original simplices, labelled by a set $S$ of participating design points. We may orient the corresponding unit vectors to have pairwise overlap $-1/r$ so that their sum vanishes and they form a regular simplex. Within any one of the original simplices that condition is already enforced, so the selected vertices at each design point can differ from their original orientations only by a common sign. Thus the selected vectors from each design point form a face of either the original simplex or its inversion.

In the inner product between vertices at different design points, only the single shared axis contributes. Thus for every pair of selected vertices from the two design points to have overlap $-1/r$, the vertices coming from a given design point must have a common sign on that axis, opposite to that of those from the other design point. Three design points having a block in common is therefore impossible, so no block can contain more than two points of $S$. Each contributing simplex thus has $u-1$ of its axes shared with other contributing simplices.

Cancellation of a given axis coordinate from the sum of vertices then requires the numbers of vertices contributed from two participating simplices to agree. Thus each design point contributes the same number $c=h/u$ of vertices to the simplex. In the rescaled $\pm1$ coordinates of Appendix~\ref{subapp:face_centroids}, the centroid of each selected face consequently has coordinates $\pm1$ on its $u-1$ shared axes and zero elsewhere.

These conditions restrict the possible values of $u$. Any omitted vertex from a contributing simplex will have inner product $-1$ with the face's centroid, and given the above characterization of the centroid, this inner product is a sum of $u-1$ signs, so $u$ must be even. If $u<h$, there will be coordinates of a contributing simplex that are not shared. Each unshared coordinate of the face's centroid is zero and is the average of $c$ signs, so $c$ must be even as well. Together with $c=h/u$, these restrictions leave the possibilities $u=2,4,14,28,56$, in addition to the case of the original simplices when $u=1$.

The $u=2$ short dependent sets are unavoidable. For any pair of design points, we can select from each either of the two $28$-vertex faces on which the shared coordinate is constant. Any choice can be oriented so that the two sums cancel, giving exactly four short dependent sets for each pair of design points.

Now suppose, as we will arrange below, that no short dependent set involves more than two design points. A pair of lines from disparate original simplices will belong to exactly one short dependent set, namely the unique two-simplex set containing them. On the other hand, a pair of lines from a single original simplex will belong to that simplex as well as to $432=27\cdot8\cdot2$ two-simplex sets: the corresponding vertices agree on $27$ of their $55$ coordinate signs, and each of those axes is shared with eight other simplices, at each of which either $28$-vertex face with constant shared coordinate can be chosen. Thus, in this case, two lines come from the same original simplex if and only if they occur together in more than one short dependent set.

This allows us to recover the original grouping into simplices from the lines alone. Consequently, any orthogonal symmetry of such a frame must permute the spans $W_x$ of the design points as well as their intersections $W_x\cap W_y=\IR e_{B(x,y)}$, \ie, the block axes. However, as we place inequivalent Hadamard simplices in each design point subspace, a nontrivial permutation of the $W_x$ spans is disallowed (see Section~\ref{sec:construction}). Thus the only possible symmetries will be a collection of reflections of block axes.

\subsection{\label{subapp:extra_simplices}Excluding additional short dependent sets}

Let $\MM_u$ be the set of signed matching choices admitting at least one short dependent set involving $u$ design points. We want to bound the size of this set for $u=4,14,28,56$. For the first three cases, it will suffice to observe that for a given set of design points, only a small fraction of matching choices at each of those points will allow for the shared axes to support the centroid of an appropriate face of the simplex.

Let $A_u$ be an upper bound on the number of distinct $(u-1)$-axis sets supporting the centroid of a $c=56/u$-vertex face of any model simplex (\ie, with centroid coordinates $\pm1$ on that set and vanishing elsewhere). Now fix a candidate set $S$ of $u$ design points. At any $x\in S$, at most $A_u$ of the $\binom{55}{u-1}$ possible subsets support a suitable face. A given set of $u-1$ coordinates is assigned to the shared axes by $2^{55}(u-1)!(55-u+1)!$ of the $2^{55}55!$ signed matchings. Thus the fraction of matchings that could conceivably give rise to a short dependent set through the points of $S$ is at most $(A_u/\binom{55}{u-1})^u$. Summing over at most $\binom{441}{u}$ choices of $S$ gives
\begin{equation}\label{eq:local_count}
    p(\MM_u)\leqslant\binom{441}{u}
       \left(\frac{A_u}{\binom{55}{u-1}}\right)^u~.
\end{equation}
This is a very coarse bound, but will already suffice for our purposes.

Consider first $u=14$ and $u=28$, in which case each face has only four or two vertices, respectively. Even allowing every face to count (although some quadruples of vertices will not actually have centroids with the right support conditions) gives
\begin{equation*}
    A_{14}\leqslant\binom{56}{4}~,
    \qquad A_{28}\leqslant\binom{56}{2}~.
\end{equation*}
These crude estimates are already more than sufficient. Indeed, substituting into \eqref{eq:local_count}, we find
\begin{equation}\label{eq:small_counts}
\begin{split}
    p(\MM_{14})&\leqslant\binom{441}{14}\left(\frac{\binom{56}{4}}{\binom{55}{13}}\right)^{14}<10^{-66}~,\\
    p(\MM_{28})&\leqslant\binom{441}{28}
       \left(\frac{\binom{56}{2}}{\binom{55}{27}}\right)^{28}<10^{-302}~.
\end{split}
\end{equation}

For $u=4$, the same estimate is insufficient: there are many more fourteen-vertex faces than triples of coordinate axes. Here we use the result of Appendix~\ref{subapp:face_centroids} that $A_4\leqslant 27$. Thus
\begin{equation}\label{eq:four_point_count}
    p(\MM_4)\leqslant\binom{441}{4}
       \left(\frac{27}{\binom{55}{3}}\right)^4<\frac{1}{573}~.
\end{equation}

Finally, for $u=56$ each point supplies one vertex and all $55$ axes are shared, so the above argument imposes no restrictions and we will use a different analysis. Fix a set $S$ of $56$ design points with no three sharing a block. Now choose one vertex at each point. On each shared axis, the images of the two vertices must have opposite signs, which holds for half of the orientation choices associated with that axis for the two points. Different pairs of points impose disjoint constraints, since no three of the points share a block. Thus there are $\binom{56}{2}=1540$ distinct sign-matching conditions.

There are at most $\binom{441}{56}$ choices of design points (without imposing the condition that no three share a block), and $112^{56}$ choices of sets of oriented vertices that could form a $u=56$ simplex from them before imposing the sign-matching constraints. Thus, the overall fraction with $u=56$ simplices obeys the bound
\begin{equation}\label{eq:large_count}
    p(\MM_{56})\leqslant\binom{441}{56}\,112^{56}\,2^{-1540}
         <10^{-277}~.
\end{equation}

From these considerations, the $u=4$ term dominates the estimates by far. The total fraction is less than $1/573+10^{-66}+10^{-277}+10^{-302}<1/572$, as used in Lemma~\ref{lem:simplex_recovery}.

\subsection{Excluding symmetries}

Now consider a potential symmetry of the frame, which in the absence of additional short dependencies will act as a sign on each block axis. We assume the symmetry is not the identity or a global inversion. We will say that for such a potential symmetry, a design point is \emph{constant} if all of its axes are either fixed or reflected. Otherwise we say a design point is \emph{mixed}. We note immediately that because any two design points share an axis, all constant points must agree on whether they are fixed or reflected.

Thus, for a symmetry other than $\pm\id$ there must be at least one mixed point. At a mixed point, in order for the symmetry to preserve the frame lines, it must act with the prescribed $27/28$ split described in Appendix~\ref{subapp:doubled_simplex}, \ie, it must reflect a specific set of $27$ axes or the complementary $28$ axes. Thus the symmetry can only be compatible with a fraction of at most $2/\binom{55}{27}$ of the possible axis matchings at that design point.

If not all points are mixed, then the constant points determine a fixed overall sign, in which case at a mixed point, there will be at least $27$ axes whose sign disagrees with the global one. Each of those axes belongs to eight other mixed design points, and these sets of eight are disjoint, so there are at least $217=1+27\cdot8$ mixed points. If there are no constant points, then there are $441$ mixed points, so the lower bound of $217$ holds in general.

There are $D=2695$ block axes and hence fewer than $2^D$ nonconstant sign choices for a possible symmetry. For each one the total proportion of matchings is bounded above by $(2/\binom{55}{27})^{217}$, so if $\MM_{\rm refl}$ is the set of signed matching choices for which some such symmetry preserves each original simplex, we have the overall bound
\begin{equation}\label{eq:reflection_count}
    p(\MM_{\rm refl})\leqslant2^{2695}
       \left(\frac{2}{\binom{55}{27}}\right)^{217}<10^{-2504}~.
\end{equation}

Combining \eqref{eq:small_counts}, \eqref{eq:four_point_count}, \eqref{eq:large_count}, and \eqref{eq:reflection_count}, the total exceptional fraction is less than $1/572$,\!\footnote{Our bounds have been extremely crude, and the true fraction of bad cases is surely much smaller than this.} and outside of that fraction, the only orthogonal symmetries are $\pm\id$, which proves \eqref{eq:bad_fraction} and the existence claim used in Section~\ref{sec:construction}.

\section{\label{app:smaller}Possible constructions with fewer fields}

\begin{table}[t]
\centering
\renewcommand{\arraystretch}{1.2}
\begin{tabular}{c|cc}
    Steiner design & $2$-$(153,9,1)$ & $2$-$(136,10,1)$\\\hline\hline
    $(r,h,D)$ & $(19,20,323)$ & $(15,16,204)$\\
    $(N,M)$ & $(2737,3060)$ & $(1972,2176)$\\
    $(\rho,\mu)$ & $(180/161,1/161)$ & $(32/29,1/145)$\\
    $B^2-4AC$ & $262800/161^4$ & $4777024/3978455625$
\end{tabular}
\caption{\label{tab:conditional}Naimark complement frame parameters for the two hypothetical Steiner designs displayed in the top row. Both satisfy the fixed point conditions \eqref{eq:window}. With a suitable choice of Hadamard matrix, we expect most signed matchings to give asymmetric frames and hence asymmetric one- and two-loop fixed points.}
\end{table}

In Table \ref{tab:conditional} we record the parameters of two potential constructions that could reduce the number of fields needed for an asymmetric one-loop fixed point. Each requires a Steiner $2$-design whose existence has not been established.

For these cases, there are not enough inequivalent Hadamard matrices of the given order to place an inequivalent Hadamard simplex at each design point; instead one will use the same model simplex at each point. For the first case ($N=2737$) one can use any Hadamard matrix of order $20$ and the same parity argument as at order $28$ rules out nontrivial coordinate-reflection symmetries of the simplex vertices. For the second case ($N=1972$) one should use the unique equivalence class (out of five) of Hadamard matrices of order $16$ whose simplex vertices admit no nontrivial coordinate-reflection symmetry (this is the matrix \texttt{had.16.4} in Sloane's library \cite{SloaneHadamard}, and the absence of such symmetries can be checked directly). We expect that counting arguments like those of Appendix~\ref{app:asymmetry}, supplemented by a bound on matchings compatible with a nontrivial permutation of the simplices, will show that a design with these parameters admits asymmetric signed matchings; we have not carried out the full analysis. Taking Naimark complements would then give equiangular tight frames with parameters as in the table.

\vfill\eject

\providecommand{\bysame}{\leavevmode\hbox to3em{\hrulefill}\thinspace}
\providecommand{\href}[2]{#2}


\begin{thebibliography}{FJKM18Z}
\bibitem[ABG04]{AbelBluskovGreig2004}
R.~Julian~R. Abel, Iliya Bluskov, and Malcolm Greig, \emph{{Balanced incomplete
  block designs with block size 9: Part III}}, Australas. J. Combin.
  \textbf{30} (2004), 57--73.

\bibitem[Aha73]{Aharony:1973zz}
Amnon Aharony, \emph{{Critical Behavior of Anisotropic Cubic Systems}}, Phys.
  Rev. B \textbf{8} (1973), 4270--4273.

\bibitem[BLZ74]{Brezin:1973jt}
E.~Br\'ezin, J.~C. Le~Guillou, and J.~Zinn-Justin, \emph{{Discussion of
  critical phenomena for general $n$-vector models}}, Phys. Rev. B \textbf{10}
  (1974), 892--900.

\bibitem[CSVZ20]{Codello:2020lta}
Alessandro Codello, Mahmoud Safari, Gian~Paolo Vacca, and Omar Zanusso,
  \emph{{Critical models with $N\leq4$ scalars in $d=4-\epsilon$}}, Phys. Rev.
  D \textbf{102} (2020), no.~6, 065017,
  \href{https://arxiv.org/abs/2008.04077}{arXiv:2008.04077}.

\bibitem[CD07]{ColbournDinitz2007}
Charles~J. Colbourn and Jeffrey~H. Dinitz (eds.), \emph{{Handbook of
  Combinatorial Designs}}, 2nd ed., Discrete Mathematics and Its Applications,
  Chapman \& Hall/CRC, Boca Raton, 2007.

\bibitem[FJKM18]{FickusJasperKingMixon2018}
Matthew Fickus, John Jasper, Emily~J. King, and Dustin~G. Mixon,
  \emph{{Equiangular tight frames that contain regular simplices}}, Linear
  Algebra Appl. \textbf{555} (2018), 98--138,
  \href{https://arxiv.org/abs/1711.07081}{arXiv:1711.07081}.

\bibitem[FMT12]{FickusMixonTremain2012}
Matthew Fickus, Dustin~G. Mixon, and Janet~C. Tremain, \emph{{Steiner
  equiangular tight frames}}, Linear Algebra Appl. \textbf{436} (2012), no.~5,
  1014--1027, \href{https://arxiv.org/abs/1009.5730}{arXiv:1009.5730}.

\bibitem[GS70]{GoethalsSeidel1970}
J.~M. Goethals and J.~J. Seidel, \emph{{Strongly regular graphs derived from
  combinatorial designs}}, Canad. J. Math. \textbf{22} (1970), no.~3,
  597--614,
  \href{https://doi.org/10.4153/CJM-1970-067-9}{doi:10.4153/CJM-1970-067-9}.

\bibitem[Gre98]{Greig1998}
Malcolm Greig, \emph{\href{https://combinatorialpress.com/jcmcc-articles/volume-027/some-group-divisible-design-constructions/}{Some group divisible design constructions}}, J. Combin.
  Math. Combin. Comput. \textbf{27} (1998), 33--52.

\bibitem[JO90]{Jack:1990eb}
I.~Jack and H.~Osborn, \emph{{Analogs of the $c$-theorem for four-dimensional
  renormalisable field theories}}, Nucl. Phys. B \textbf{343} (1990),
  647--688.

\bibitem[Kim94]{Kimura1994}
Hiroshi~Kimura, \emph{{Classification of {H}adamard matrices of order 28}},
Discrete Math. \textbf{133} (1994), no.~1--3, 171--180.

\bibitem[LS73]{LemmensSeidel1973}
P.~W.~H. Lemmens and J.~J. Seidel, \emph{{Equiangular lines}}, J. Algebra
  \textbf{24} (1973), no.~3, 494--512.

\bibitem[Mic84]{Michel:1983in}
Louis Michel, \emph{{Renormalization-group fixed points of general $n$-vector
  models}}, Phys. Rev. B \textbf{29} (1984), 2777--2783.

\bibitem[OS18]{Osborn:2017ucf}
Hugh Osborn and Andreas Stergiou, \emph{{Seeking fixed points in multiple
  coupling scalar theories in the $\epsilon$ expansion}}, JHEP \textbf{05}
  (2018), 051, \href{https://arxiv.org/abs/1707.06165}{arXiv:1707.06165}.

\bibitem[OS21]{Osborn:2020cnf}
\bysame, \emph{{Heavy handed quest for fixed points in multiple coupling scalar
  theories in the $\epsilon$ expansion}}, JHEP \textbf{04} (2021), 128,
  \href{https://arxiv.org/abs/2010.15915}{arXiv:2010.15915}.

\bibitem[RR24]{Rong:2023xhz}
Junchen Rong and Slava Rychkov, \emph{{Classifying irreducible fixed points of
  five scalar fields in perturbation theory}}, SciPost Phys. \textbf{16}
  (2024), no.~2, 040, \href{https://arxiv.org/abs/2306.09419}{arXiv:2306.09419}.

\bibitem[RS19]{Rychkov:2018vya}
Slava Rychkov and Andreas Stergiou, \emph{{General Properties of Multiscalar RG
  Flows in $d=4-\varepsilon$}}, SciPost Phys. \textbf{6} (2019), no.~1, 008,
  \href{https://arxiv.org/abs/1810.10541}{arXiv:1810.10541}.

\bibitem[Slo]{SloaneHadamard}
N.~J.~A. Sloane, \emph{{A Library of Hadamard Matrices}},
  \url{http://neilsloane.com/hadamard/}, accessed 15 October 2026.

\bibitem[STDH07]{sustik2007existence}
M.~A. Sustik, J.~A. Tropp, I.~S. Dhillon, and R.~W. Heath, Jr.,
  \emph{{On the existence of equiangular tight frames}}, Linear Algebra Appl.
  \textbf{426} (2007), no.~2--3, 619--635.

\bibitem[Wal18]{Waldron2018}
Shayne F.~D. Waldron, \emph{{An Introduction to Finite Tight Frames}}, Applied
  and Numerical Harmonic Analysis, Birkh{\"a}user, New York, 2018.

\bibitem[WF72]{Wilson:1971dc}
Kenneth~G. Wilson and Michael~E. Fisher, \emph{{Critical exponents in 3.99
  dimensions}}, Phys. Rev. Lett. \textbf{28} (1972), 240--243.

\bibitem[ZW75]{Zia:1975ha}
R.~K.~P. Zia and D.~J. Wallace, \emph{{Critical behaviour of the continuous
  $n$-component Potts model}}, J. Phys. A \textbf{8} (1975), 1495--1507.

\end{thebibliography}
\end{document}